\documentclass[aps,pra,superscriptaddress,twocolumn,shortbibliography]{revtex4-1}
\usepackage{mathrsfs}
\usepackage[english]{babel}
\usepackage{amsfonts}
\usepackage{amssymb}
\usepackage{amsmath}
\usepackage{graphicx}
\graphicspath{{figures/}{./}} 
\usepackage{sidecap}
\usepackage{soul}
\usepackage{color}
\usepackage[colorlinks=true, urlcolor=blue, citecolor=blue,linkcolor=blue,citebordercolor={1 0 0},linkbordercolor={0 0 1}]{hyperref}
\usepackage{subeqnarray}
\usepackage{verbatim}
\usepackage{euscript} 
\usepackage{lipsum}
\usepackage{hyperref}
\usepackage{bibunits}
\usepackage{algorithm}
\usepackage{algpseudocode}
\usepackage{appendix}
\usepackage{quantikz}
\usepackage{dsfont}
\usepackage{txfonts}
\usepackage{array}
\usepackage{multirow}
\usepackage[normalem]{ulem}

\makeatletter
\renewcommand\frontmatter@abstractwidth{\dimexpr\textwidth -0in \relax}
\makeatother
\renewcommand{\eqref}[1]{Eq.~(\ref{#1})} 

\begin{document}

\title{Stable many-body resonances in open quantum systems} 

\author{Ruben Pe\~{n}a}
\affiliation{Departamento de F\'isica, Universidad de Santiago de Chile, 
Avenida V\'ictor Jara 3493, 9170124, Santiago, Chile}

\author{Thi Ha Kyaw}
\affiliation{LG Electronics Toronto AI Lab, Toronto, Ontario M5V 1M3, Canada}

\author{Guillermo Romero}
\email{guillermo.romero@usach.cl}
\affiliation{Departamento de F\'isica, CEDENNA, Universidad de Santiago de Chile, 
Avenida V\'ictor Jara 3493, 9170124, Santiago, Chile}
\
\date{\today}

\begin{abstract}
Periodically driven quantum many-body systems exhibit novel nonequilibrium states such as prethermalization, discrete time crystals, and many-body localization. Recently, the general mechanism of fractional resonances has been proposed that leads to slowing the many-body dynamics in systems with both $U(1)$ and parity symmetry. Here, we show that fractional resonance is stable under local noise models. To corroborate our finding, we numerically study the dynamics of a small-scale Bose-Hubbard model that can readily be implemented in existing noisy intermediate-scale quantum (NISQ) devices. Our findings suggest a possible pathway towards a stable nonequilibrium state-of-matter, with potential applications of quantum memories for quantum information processing.
\end{abstract}
\maketitle

\section{Introduction}
Nonequilibrium quantum phases without static analogs have become an active area of research since programmable quantum simulators \cite{PRXQuantum.2.017003} such as cold atoms \cite{RevModPhys.80.885,Cheneau:2012aa,Bernien:2017aa,Choi1547} and superconducting circuits \cite{Roushan1175,Ma:2019aa,PhysRevLett.123.050502,PhysRevLett.125.170503,PhysRevResearch.3.033043,Neill195} allow the preparation of exotic states of matter into out-of-equilibrium. Paradigmatic examples of nonequilibrium states are discrete time crystals \cite{PhysRevA.91.033617,PhysRevLett.117.090402,PhysRevLett.118.030401,Sacha_2017,PhysRevLett.123.150601,TimeCrystals,Pizzi:2021we,PhysRevLett.127.140602,PhysRevB.104.094308,PhysRevResearch.3.L042023} and Floquet prethermalization \cite{PhysRevLett.115.256803,PhysRevB.95.014112,Abanin:2017uy,PhysRevLett.116.120401,KUWAHARA201696,PhysRevX.10.021044,PhysRevA.105.012418,torre2021statistical}. The search for nonequilibrium states of matter is challenging since one needs to move beyond standard quantum statistical mechanics \cite{Dziarmaga2010,RevModPhys.83.863,Eisert:2015aa,Mitra_Quench,Heyl_2018}, and simulations of quantum many-body systems in classical computers \cite{Orus:2019aa,SCHOLLWOCK201196,10.21468/SciPostPhysLectNotes.8}. In a recent contribution by some of the authors \cite{PhysRevB.106.064307}, it has been pointed out the emergence of a prethermal and localized nonequilibrium phase, termed ``fractional resonance", appearing in a broad class of many-body Hamiltonians exhibiting U(1) and parity symmetry such as the Bose-Hubbard model~\cite{PhysRevB.40.546,Jaksch1998}, the $XXZ$ spin-1 model \cite{PhysRevB.67.104401,PhysRevLett.126.163203} or the Jaynes-Cummings-Hubbard model \cite{Greentree:2006aa,Hartmann:2006aa,PhysRevA.76.031805}. 
Moreover, small-scale versions of these models can be experimentally implemented in NISQ devices. Therefore, we believe it is pertinent to investigate for the stability of such nonequilibrium states of matter under the action of loss mechanisms, which will be relevant for applications in quantum memories~\cite{doi:10.1073/pnas.1819316116} and quantum metrology~\cite{PhysRevLett.127.080504}.

On the other hand, quantum computing~\cite{Nielsen2010quantum} and quantum machine learning~\cite{Wittek2014quantum,Carleo2019machine,Schuld2021machine} are progressing tremendously. The main driving force behind this progress is that quantum computers promise exponential speedup over their classical counterparts in solving specific tasks \cite{Shor1994algorithms}. When fault-tolerant general-purpose quantum computers become available in the long term, it is expected to implement adiabatic state preparation and quantum phase estimation as the standard quantum routine to determine the ground state energy of a sophisticated physical Hamiltonian~\cite{Aspuru-Guzik2005simulated,Georgescu2014quantum,Cao2019quantum,McArdle2020quantum}.
However, such schemes are resource-intensive, and thus they are not appropriate for current NISQ hardware~\cite{Preskill2018quantum,Arute2019quantum,Zhong2020quantum,Wu2021strong,Madsen2022quantum}, thereby shifting the central research theme towards low-depth hybrid quantum-classical algorithms, otherwise known as NISQ algorithms \cite{Bharti2022noisy}. Recently, by utilizing quantum data processing based on the classical shadow, learning certain information about a many-body quantum state from an experiment can be exponentially reduced~\cite{Chen2021exponential,Huang2021quantum,Aharonov2022quantum}.
In contrast, in the standard classical paradigm, one must repeat the experiment many times to build statistical certainty about the measurement data at each experimental run. By loading the entire quantum state of a many-body system into a quantum memory, a shift in the paradigm occurs in a new approach. After some quantum processing is applied to the replicated state, measurements are performed. Then the protocol can be repeated. Furthermore, renowned quantum machine learning algorithms center around the Harrow-Hassidim-Lloyd (HHL) algorithm \cite{Harrow2009quantum} and quantum principal component analysis \cite{Lloyd2014quantum}, which require the physical realization of quantum random access memory~\cite{Giovannetti2008quantum,Kyaw2015scalable}. Here we recognize a substantial overlap with localized nonequilibrium states of matter, such as prethermal states \cite{PhysRevLett.115.256803,PhysRevB.95.014112,Abanin:2017uy,PhysRevLett.116.120401,KUWAHARA201696,PhysRevX.10.021044,PhysRevA.105.012418,torre2021statistical} and many-body localization~\cite{Huse2015,RevModPhys.91.021001,ALET2018498}. Moreover, because localized nonequilibrium states of matter retain information about their initial state, they may be used as quantum memories~\cite{PhysRevB.105.205133}. These and the above arguments indicate that developing and realizing a quantum memory represents the pinnacle achievement in quantum technologies and applications.

In this article, we provide a detailed analysis of the stability of fractional many-body resonances under noisy environments within the paradigm of NISQ devices. This investigation represents an essential step before seeking potential applications of fractional resonances and their associated prethermal states as a quantum memory. The article is organized as follows. In section \ref{model}, we briefly recap the emergence of many-body resonances, emphasizing the integer and fractional resonance states using a generic Hamiltonian. In section \ref{opendynamics}, we consider a one-dimensional lattice of strongly correlated bosonic particles described by the Bose-Hubbard model (BHM) \cite{PhysRevB.40.546,Jaksch1998}, where we study the open quantum system dynamics of the BHM, and discuss the stability of integer and fractional resonance states when considering realistic parameters of NISQ devices such as superconducting circuits, for the sake of simulating the prescribed physics in an experiment. Finally, in section \ref{conclusion}, we present our concluding remarks.

\section{The model}
\label{model}
Let us consider a one-dimensional lattice with open boundary conditions described by the generic lattice Hamiltonian
\begin{equation}
\label{eq:HamAlgebra}
\hat{H}=\hbar\sum^L_{j=1}\big(\omega \hat{O}_j+\frac{U}{2}\hat{O}^2_j\big)-\hbar J_0\cos{(\Omega t)}\sum^{L-1}_{j=1}\big(\hat{A}^{\dagger}_j\hat{A}_{j+1}+\hat{A}^{\dagger}_{j+1}\hat{A}_j\big),
\end{equation}
which is composed of a local energy term $\hat{H}_0\equiv\hbar\sum_{j=1}^L\big(\omega\hat{O}_j+U\hat{O}^2_j/2\big)$ with local operators $\hat{O}_j$, and a time-dependent hopping interaction $\hat{H}_I(t)=\hbar J_0\cos{(\Omega t)}\sum^{L-1}_{j=1}\big(\hat{A}^{\dagger}_j\hat{A}_{j+1}+\hat{A}^{\dagger}_{j+1}\hat{A}_j\big)$, allowing for the exchange of particles/excitations between the nearest neighboring sites via the local ladder operators $\hat{A}_j$ and $\hat{A}^{\dagger}_j$. These operators satisfy the commutation relations $[\hat{O}_i,\hat{A}^{\dagger}_j]=\delta_{ij}\hat{A}^{\dagger}_j$, $[\hat{O}_i,\hat{A}_j]=-\delta_{ij}\hat{A}_j$. 
Here, we restrict ourselves to studying bosonic particles.
However, the treatment is also applicable to spin-$d$ systems.
If the many-body system is isolated from an external environment, the term $\sum^N_{j=1}\omega \hat{O}_j$ is a constant of motion since the Hamiltonian exhibits U(1) symmetry. The Hermitian operator $\hat{O}_j$ satisfies the eigenvalue equation $\hat{O}_j\ket{m_j}=m_j\ket{m_j}$, where $m_j$ is a quantum number that labels the local quantum state of the $j$th lattice site. For instance, $m_j$ may represent the occupation number of a bosonic system or the spin component along the $z$ direction of spin-$d$ systems. In this work, we focus on finite-size lattices with reflection symmetry characterized by the parity operator $\mathcal{\hat{P}}$, which satisfies  $\mathcal{\hat{P}}\ket{m_1,\hdots,m_j,\hdots,m_L}=\ket{m_L,\hdots,m_j,\hdots,m_1}$, where $|m_1,..,m_j,..,m_L\rangle\equiv\bigotimes_{l=1}^L|m_l\rangle$.

To understand the processes that may occur due to the hopping of particles/excitations, let us move to a rotating frame with respect to the free Hamiltonian $\hat{H}_0$. We obtain
\begin{equation}
\begin{aligned}
\widetilde{H}_{I}(t)=&-\hbar J_0\cos{(\Omega t)}\sum^{L-1}_{j=1}\Big[e^{iU t\big(\hat{O}_{j+1}-\hat{O}_{j}-1\big)}\hat{A}^{\dagger}_j\hat{A}_{j+1}\\
&+\hat{A}^{\dagger}_{j+1}\hat{A}_j e^{-iU t\big(\hat{O}_{j+1}-\hat{O}_{j}-1\big)}\Big],
\label{eq:OperatorEv}
\end{aligned}
\end{equation}
from which we can identify two characteristic time scales: one is provided by the driving frequency $\Omega$, and another is the on-site interaction $U$, which leads to a local anharmonic spectrum. The Hamiltonian~(\ref{eq:OperatorEv}) is not strictly periodic either in $\Omega$ or $U$. However, we will consider integer and fractional driving frequencies in the unit of $U$ defined as $\Omega=U$ and $\Omega=U/2$, respectively. In both cases, it can be shown that $\widetilde{H}_I(t+T)=\widetilde{H}_I(t)$ is periodic with period $T=2\pi/\Omega$~\cite{PhysRevB.106.064307}. Also, we consider the strongly interacting regime where $U\gg J_0$ \cite{Cheneau:2012aa}, where $J_0$ is the bare (static) hopping rate. This regime allows us to describe the particle/excitation hopping processes within the semi-classical picture as discussed in the Refs.~\cite{torre2021statistical,PhysRevB.106.064307}.

Let us briefly describe the aforementioned semi-classical picture to identify many-body resonances \cite{CHIRIKOV1979263} in the closed system scenario. To find many-body resonances, we focus on how specific quantum states, referred to as configurations, are connected via the hopping term. Let us suppose the many-body system is initialized in the configuration $\ket{m_1,\hdots,m_j,m_k,m_l,\hdots,m_L}$, a hopping event from the site $j$ to $k$ will lead the system to occupy the state $\ket{m_1,\hdots,m_j-1,m_k+1,m_l,\hdots,m_L}$. Now, let us compute the local energy of those states using the unperturbed Hamiltonian $\hat{H}_0$. The result reads
\begin{widetext}
\centering
\begin{subequations}
\begin{align}
&\hat{H}_{0}\ket{m_1,\hdots,m_j,m_k,m_l,\hdots,m_L}=\hbar\sum^L_{j=1}\left(\omega m_j+\frac{U}{2}m^2_j\right)\ket{m_1,\hdots,m_j,m_k,m_l,\hdots,m_L}\label{eq:Integer1}\\
&\hat{H}_{0}\ket{m_1,\hdots,m_j-1,m_k+1,m_l,\hdots,m_L}=\hbar\bigg[\sum^L_{j=1}\left(\omega m_j+\frac{U}{2}m^2_j\right) + U(m_k -m_j+1)\bigg]\ket{m_1,\hdots,m_j-1,m_k+1,m_l,\hdots,m_L}.
\label{matrixFRAC2}
\end{align}
\end{subequations}
\end{widetext}

The energy difference between these two configurations is $\Delta E= \hbar U(m_k -m_j+1)$. Therefore, to evolve from the state $\ket{m_1,\hdots,m_j,m_k,m_l,\hdots,m_L}$ to $\ket{m_1,\hdots,m_j-1,m_k+1,m_l,\hdots,m_L}$, the driving frequency should match the condition $m\Omega=U(m_k -m_j+1)$. Analogously, it can be shown that a hopping event from $k\to j$ leads to the condition $m\Omega=U(m_j-m_k+1)$, so we can write generically $m\Omega=U[\pm(m_k-m_j)+1]$, with $m\in\mathbb{Z}$, which defines the integer many-body resonance \cite{torre2021statistical,PhysRevB.106.064307}. In this case, the emerging many-body dynamics is ruled by nearest-neighbor interactions where the time scale for spreading particles/excitations is $J_0^{-1}$.
In our recent work \cite{PhysRevB.106.064307}, the emergence of fractional resonances in many-body quantum systems has been proposed where second-order processes rule the nonequilibrium dynamics. 

In this article, we have rigorously demonstrated the emergence of fractional resonance condition using the Magnus expansion \cite{Floquet1,Floquet2,Blanes_2010}. Here, we will follow the semi-classical picture described above to find the fractional resonance condition. 
If the initial configuration is $\ket{m_1,\hdots,m_j,m_k,m_l,\hdots,m_L}$, we want to connect this state with  $\ket{m_1,\hdots,m_j-1,m_k,m_l+1,\hdots,m_L}$ via two hopping events. Notice that the latter configuration has local energy  $\hbar\sum^L_{j=1}\left(\omega m_j+\frac{U}{2}m^2_j\right) + \hbar U(m_l -m_j+1)$. The energy difference between this configuration and the initial one is $\Delta E= \hbar U(m_l-m_j+1)$. Therefore, to connect both configurations, two identical hopping events are necessary, each kicking the system with energy $\hbar U(m_l-m_j+1)/2$. The same analysis can be done for a hopping from $j$ $\to$ $l$ leading to the condition $\Omega= U(m_j-m_l+1)/2$, so we can write generically $m\Omega=U[\pm(m_j-m_l)+1]/2$, where $m\in\mathbb{Z}$. 
\begin{figure}[t]
\centering
\includegraphics[scale=0.4]{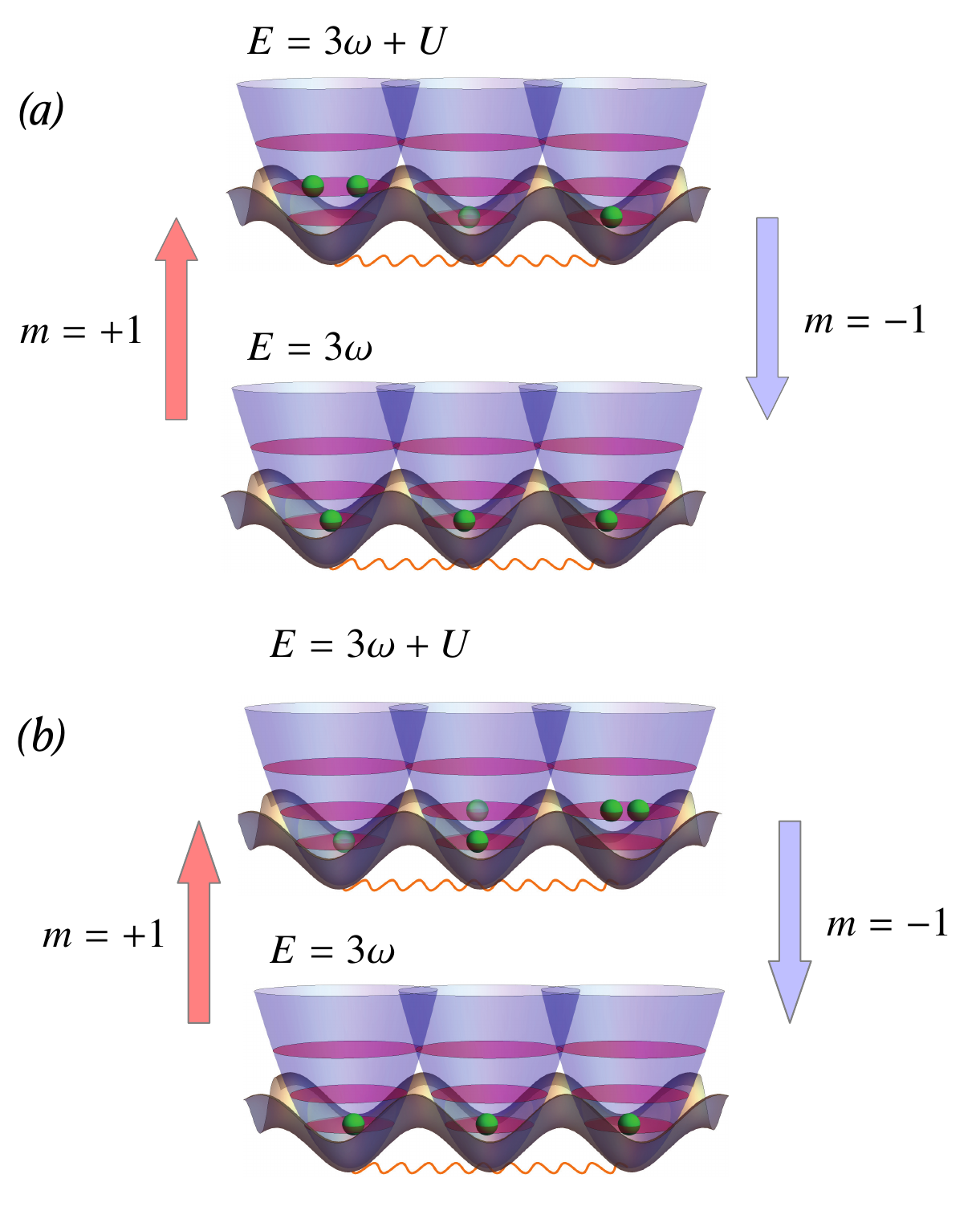}
\caption{Schematic representation of many-body resonances in the driven lattice model. a) The panel shows a transition from the initial condition $m_j=m_k=m_l=1$ with particles (green circles) occupying each lattice site (from bottom to top). Here, we represent a hopping event where the middle particle moves to the left-most site, reaching the configuration $m_j=2,m_k=0,m_l=1$. The energy difference between these configurations is $\Delta E=\hbar U$. b) The panel shows a transition from the initial condition $m_j=m_k=m_l=1$ with particles (green circles) occupying each lattice site (from bottom to top). The top panel shows two hopping events where the left-most particle moves to the right-most site, reaching the configuration $m_j=0,m_k=1,m_l=2$. The energy difference between these configurations is also $\Delta E=\hbar U$. Therefore, within the semi-classical picture, each hopping event increases the system energy by $\hbar U/2$.}
\label{Fig1}
\end{figure}

Let us consider the schematic representation of figure \ref{Fig1} to illustrate integer and fractional resonance conditions better. Here, we display a schematic representation of many-body resonances in the driven lattice model. In figure \ref{Fig1}(a), we show two configurations with all particles (green circles) occupying a lattice site and the left-most site occupied by two particles. The energy difference between these configurations is $\Delta E = \hbar U$. Therefore, to evolve from the lower to the upper configuration, one needs to activate a hopping event with frequency $\Omega = U$ to increase the system energy. In figure \ref{Fig1}(b), we show configurations that involve two hopping events where the left-most particle moves to the right-most lattice site. The energy difference between these configurations is $\Delta E = \hbar U$, so each hopping event increases the system energy by a $\hbar U/2$.

Along this work, we will prove that the emergent many-body fractional resonance is a stable phenomenon under loss mechanisms inherent to NISQ devices. In particular, we will numerically study the nonequilibrium dynamics of the Bose-Hubbard model \cite{Bose-Hubbard,Jaksch1998,PhysRevB.61.12474}, which allows us to describe implementations of strongly interacting lattice systems in superconducting circuits \cite{Roushan1175,Ma:2019aa,PhysRevLett.123.050502,PhysRevResearch.3.033043,PhysRevX.4.031043,PhysRevX.6.021044,PhysRevX.7.011016}.

\subsection{The Bose-Hubbard Model}
The Bose-Hubbard model (BHM) describes strongly interacting bosonic systems on a lattice, where we recognize operators $\hat{O}_{j}=\hat{n}_j=\hat{a}^{\dagger}_j\hat{a}_j$, $\hat{a}^{\dagger}_j=\hat{A}^{\dagger}_{j}$, and $\hat{a}_j=\hat{A}_{j}$. Replacing these operators in the generic Hamiltonian (\ref{eq:HamAlgebra}), we can generate the Bose-Hubbard Hamiltonian 
\begin{equation}
\label{eq:HBH}
\hat{H}_{\rm{BH}}(t)=\hbar\sum^L_{j=1}\big(\omega \hat{a}^{\dag}_j\hat{a}_j+\frac{U}{2}\hat{a}^{\dag}_j\hat{a}^{\dag}_j\hat{a}_j\hat{a}_j\big)-\hbar J(t)\sum^{L-1}_{j=1}(\hat{a}^{\dagger}_j\hat{a}_{j+1}+{\rm H.c.}),
\end{equation}
where $\hat{a}_j$($\hat{a}^{\dag}_j$) is the annihilation (creation) bosonic operator at site $j$th, $\omega$ is the single site frequency, $U$ the on-site interaction, $J(t) = J_0 \cos{(\Omega t)}$ the modulated hopping strength between neighboring sites, whereas $J_0$ is the bare hopping rate, and $\Omega$ the driving frequency. We stress that modulating the hopping rate can be achieved in superconducting circuits with transmons \cite{Neill195}. The symmetries of the Hamiltonian (\ref{eq:HBH}) play a crucial role in describing the many-body dynamics. In particular, the BHM exhibits $U(1)$ symmetry characterized by the conservation of the total number of particles/excitations $[\hat{H}_{\rm{BH}},\hat{N}]=0$ with $\hat{N}=\sum_{j=1}^{L}\hat{a}^{\dag}_j\hat{a}_j$. Also, since we consider a lattice with open boundary conditions, the model preserves the parity $[\hat{H}_{\rm{BH}},\hat{P}]$, where $\hat{P}\ket{n_1,n_2,\hdots,n_L}=\ket{n_L,\hdots,n_2,n_1}$, where $n_j$ stands for the number of particles/excitations at the $j$th lattice site. Hereafter, we will consider an initial state with unit filling, that is, $\ket{\psi_0}=\bigotimes_{j=1}^{L}\ket{1}_j$. In general, the total number of states that may be involved in the dynamics correspond to all possible configurations of $N$ particles distributed in $L$ lattice sites $D_{N,L}=(N+L-1)!/N!(L-1)!$.
\begin{figure}[t]
\centering
\includegraphics[scale=0.18]{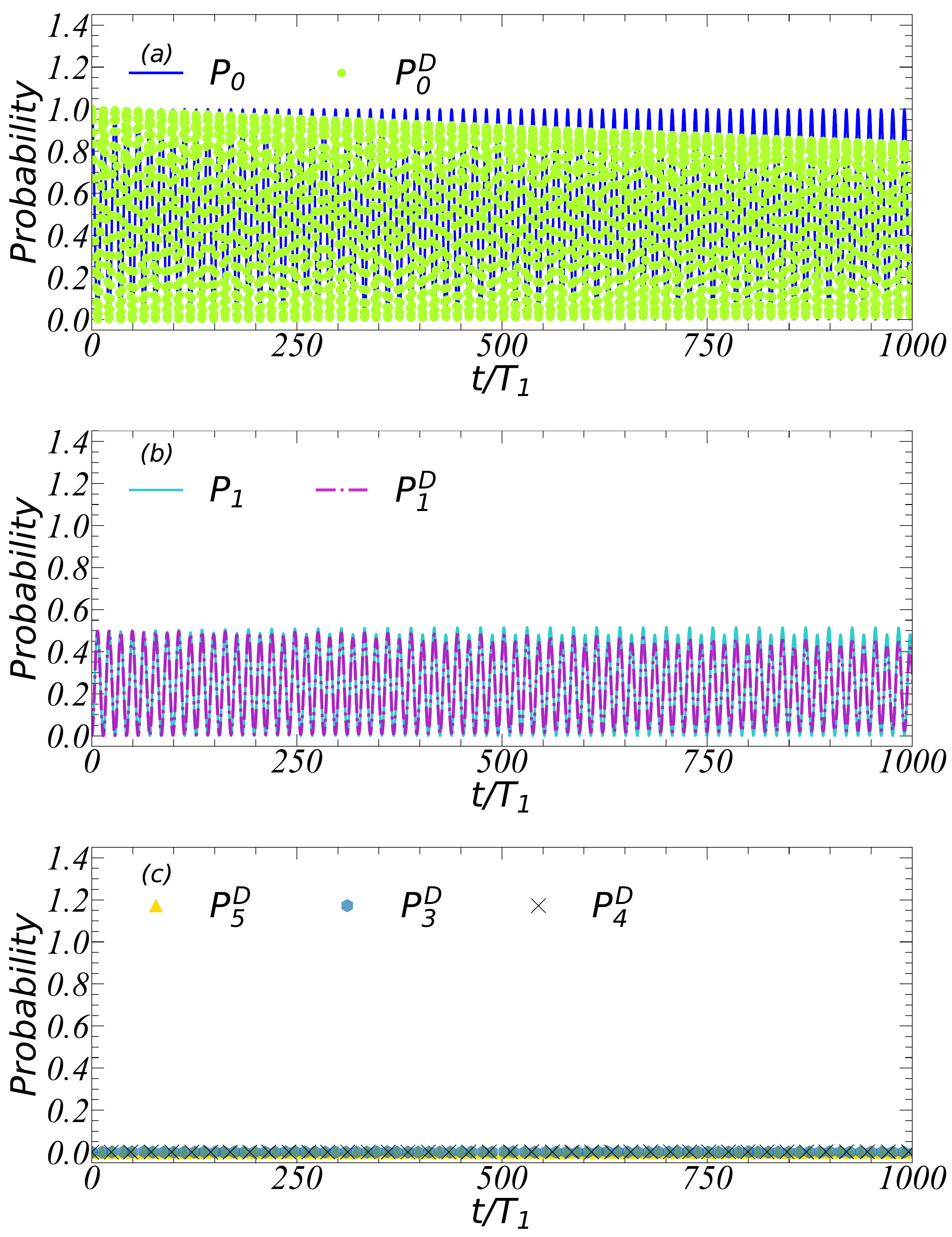}
\caption{We plot the populations of states $|\psi_i\rangle$ with $i= 0,1,2,3,4,5$ for the integer resonance $\Omega_1=U$. In all panels, $P^D_i$ are populations numerically obtained considering decay and dephasing for each lattice site, whereas $P_i$ are populations numerically computed considering a closed system scenario. Also, the three-site Bose-Hubbard lattice is initialized in the state $\ket{\psi_0}=\ket{111}$, and we use realistic parameters  $\omega=2\pi \times 4.5$ GHz, $J_0=2\pi \times 11.5$ MHz, $\kappa_{10}=11.9$ kHz, $\kappa_{21}=24.39$ kHz, $\kappa_{32}=33.33$ kHz, $\gamma_{00}=13.89$ kHz, $\gamma_{11}=31.25$ kHz, $\gamma_{22}=83.33$ kHz. We consider up to $n_{\rm max}=3$ particles per site with a local Hilbert space dimension $\dim{\mathcal{H}_\ell}=4$.}
\label{Fig2}
\end{figure}

\begin{figure}[t]
\centering
\includegraphics[scale=0.18]{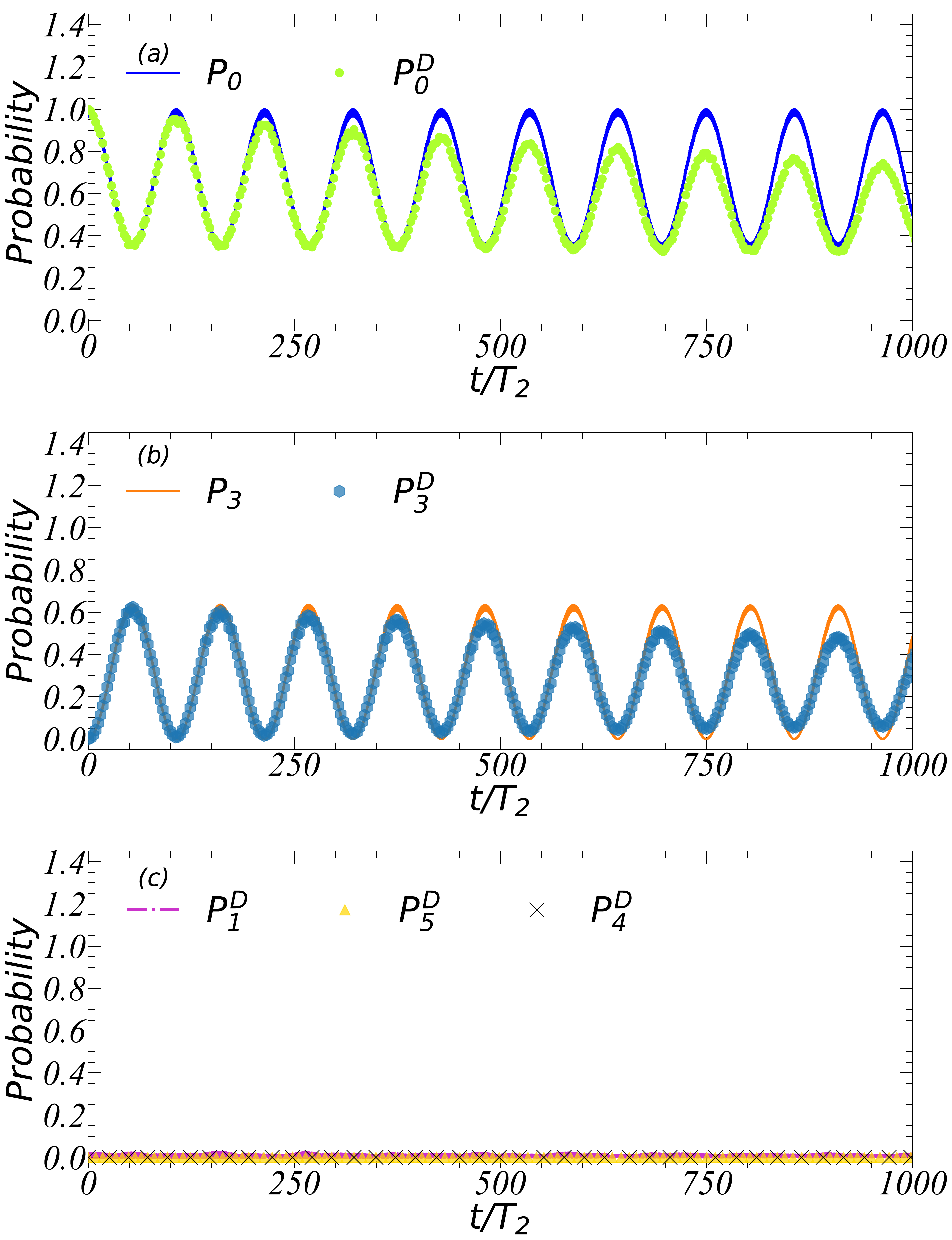}
\caption{We plot the populations of states $|\psi_i\rangle$ with $i= 0,1,2,3,4,5$ for the fractional resonance $\Omega_2=U/2$. In all panels, $P^D_i$ are populations numerically obtained considering decay and dephasing for each lattice site, whereas $P_i$ are populations numerically computed considering a closed system scenario. Also, the three-site Bose-Hubbard lattice is initialized in the state $\ket{\psi_0}=\ket{111}$, and we use realistic parameters  $\omega=2\pi \times 4.5$ GHz, $J_0=2\pi \times 11.5$ MHz, $\kappa_{10}=11.9$ kHz, $\kappa_{21}=24.39$ kHz, $\kappa_{32}=33.33$ kHz, $\gamma_{00}=13.89$ kHz, $\gamma_{11}=31.25$ kHz, $\gamma_{22}=83.33$ kHz. We consider up to $n_{\rm max}=3$ particles per site with a local Hilbert space dimension $\dim{\mathcal{H}_\ell}=4$.}
\label{Fig3}
\end{figure}

\section{Open quantum system dynamics}
\label{opendynamics}
Practical implementations of strongly correlated bosonic systems in superconducting circuits always involve interacting with some electromagnetic environment which leads to noisy dynamics. NISQ devices made of superconducting circuits have proven successful in stabilizing nonequilibrium many-body states \cite{Roushan1175,Ma:2019aa,PhysRevLett.123.050502,PhysRevLett.125.170503,PhysRevResearch.3.033043,Neill195,PhysRevA.105.012418}, whose features are well captured by the BHM \cite{PhysRevB.40.546,Jaksch1998}. In these experiments, the open system dynamics is well described by the Lindblad master equation. However, since superconducting circuits such as transmons are anharmonic oscillators, we must use the microscopic master equation to fully capture the colored nature of the bath \cite{PetruccioneBook}. Considering a single anharmonic oscillator located at the $j$th lattice site, and described by the local Hamiltonian $\hat{H}^{(j)}_S=\hbar\sum_n\omega^{(j)}_n|{n_j}\rangle\langle n_j|$, the master equation in the Lindblad form reads
\begin{equation}
\begin{aligned}
\frac{d\hat{\rho}_j}{dt} =& -\frac{i}{\hbar}[\hat{H}^{(j)}_S,\hat{\rho}_j]\\ 
 + &\sum_{m,n>m}\kappa^{(j)}_{nm}\mathbb{L}[|m_j\rangle\langle n_j|]\hat{\rho}_j+\sum_{n}\gamma^{(j)}_{nn}\mathbb{L}[|n_j\rangle\langle n_j|]\hat{\rho}_j,  
\label{OQS}
\end{aligned}
\end{equation}
where $\mathbb{L}[\hat{\mathcal{O}}]\hat{\rho}=\hat{\mathcal{O}}\hat{\rho}\hat{\mathcal{O}^{\dag}}-\frac{1}{2}(\hat{\mathcal{O}}^{\dag}\hat{\mathcal{O}}\rho+\rho\hat{\mathcal{O}}^{\dag}\hat{\mathcal{O}})$. Here, $\kappa^{(j)}_{nm}$ and $\gamma^{(j)}_{nn}$ define decay and dephasing rates, respectively. In the decay rates, the subscripts $n,m$ refer to the decay from the state $|n_j\rangle$ to $|m_j\rangle$, whereas the dephasing rates refer to the dephasing for the superpositions of states $|m_j\rangle$ and $|n_j\rangle$, within each lattice site $j$. In this work, we consider finite lattices with $L=3$ and $L=4$ sites with up to $n_{\rm{max}}=3$ and $n_{\rm{max}}=4$  particles per site with local Hilbert space dimension $\rm{dim}(\mathcal{H}_\ell)=4$ and $\rm{dim}(\mathcal{H}_\ell)=5$, respectively. In these cases, a multi-level approach with local states $\{|0_j\rangle,|1_j\rangle,|2_j\rangle,|3_j\rangle\}$ and $\{|0_j\rangle,|1_j\rangle,|2_j\rangle,|3_j\rangle,|4_j\rangle\}$ must be included in the Lindblad master equation (\ref{OQS}).

Using the multi-level approach for a single anharmonic oscillator, in this work, we consider the Bose-Hubbard lattice described by the master equation
\begin{align}
\frac{d\hat{\rho}}{dt}&=-\frac{i}{\hbar}[\hat{H}_{\rm BH}(t),\hat{\rho}]+\sum_{j=1}^L\sum_{m=0,n>m}^{n_{\textrm{max}}}\kappa^{(j)}_{nm}\mathbb{L}[|m_j\rangle\langle n_j|]\hat{\rho}\nonumber\\
&+ \sum_{j=1}^L\sum_{n=0}^{n_{\textrm{max}}}\gamma^{(j)}_{nn}\mathbb{L}[|n_j\rangle\langle n_j|]\hat{\rho},
\label{MEBH}
\end{align}
where $\hat{\rho}$ represents the density matrix of the strongly interacting lattice.

\begin{figure}[t]
\centering
\includegraphics[scale=0.18]{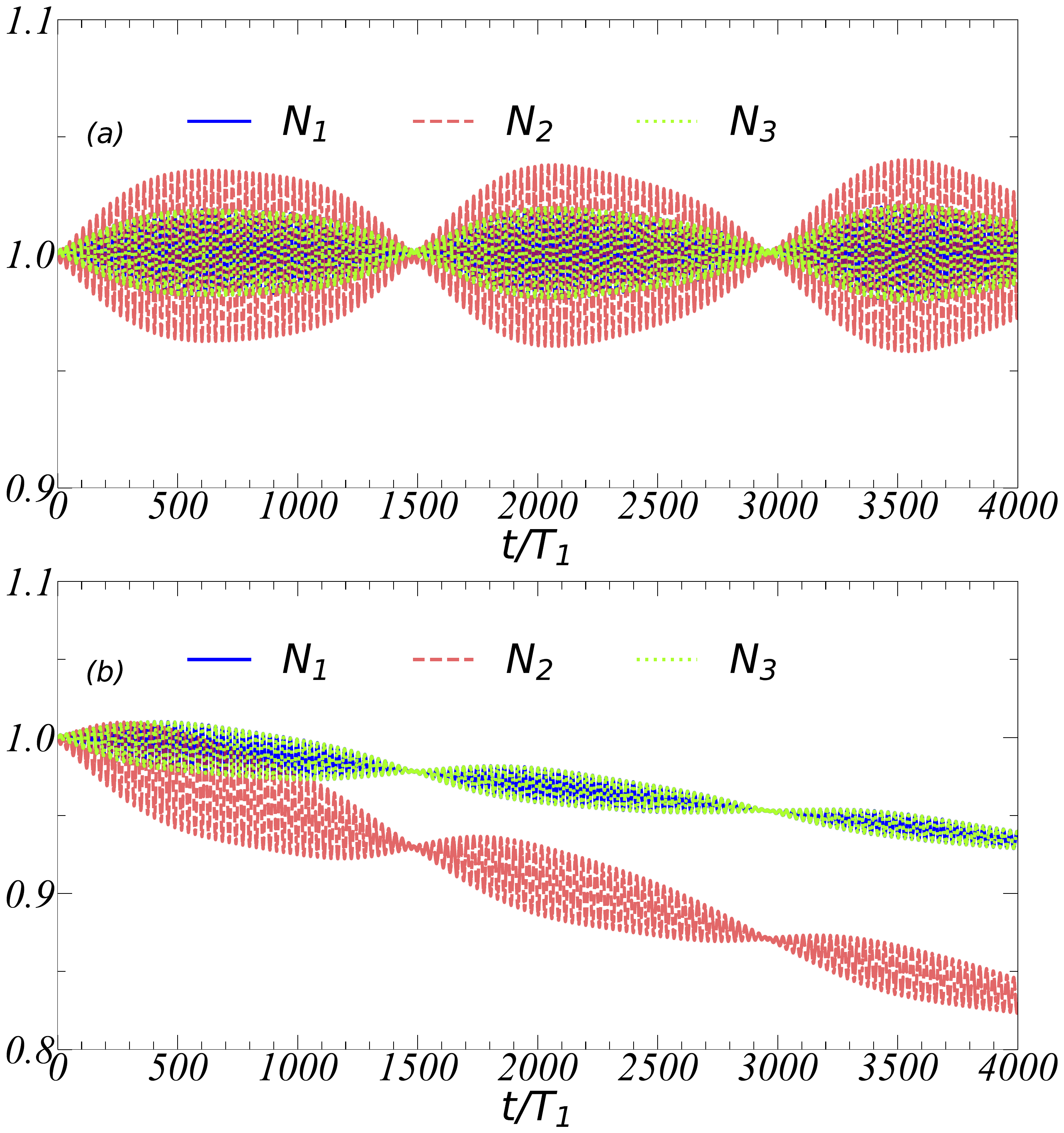}
\caption{We plot the average occupation number $\langle N_j(t) \rangle$ for each lattice site, for the integer resonance $\Omega_1=U$. The upper panel (a) shows the closed system scenario. The lower panel (b) shows the open system scenario. As in previous numerical simulations, the three-site Bose-Hubbard lattice is initialized in the state $\ket{\psi_0}=\ket{111}$. We use realistic parameters  $\omega=2\pi \times 4.5$ GHz, $J_0=2\pi \times 11.5$ MHz, $\kappa_{10}=11.9$ kHz, $\kappa_{21}=24.39$ kHz, $\kappa_{32}=33.33$ kHz, $\gamma_{00}=13.89$ kHz, $\gamma_{11}=31.25$ kHz, $\gamma_{22}=83.33$ kHz. We consider up to $n_{\rm max}=3$ particles per site with a local Hilbert space dimension $\dim{\mathcal{H}_\ell}=4$.}
\label{Fig4}
\end{figure}

\begin{figure}[t]
\centering
\includegraphics[scale=0.18]{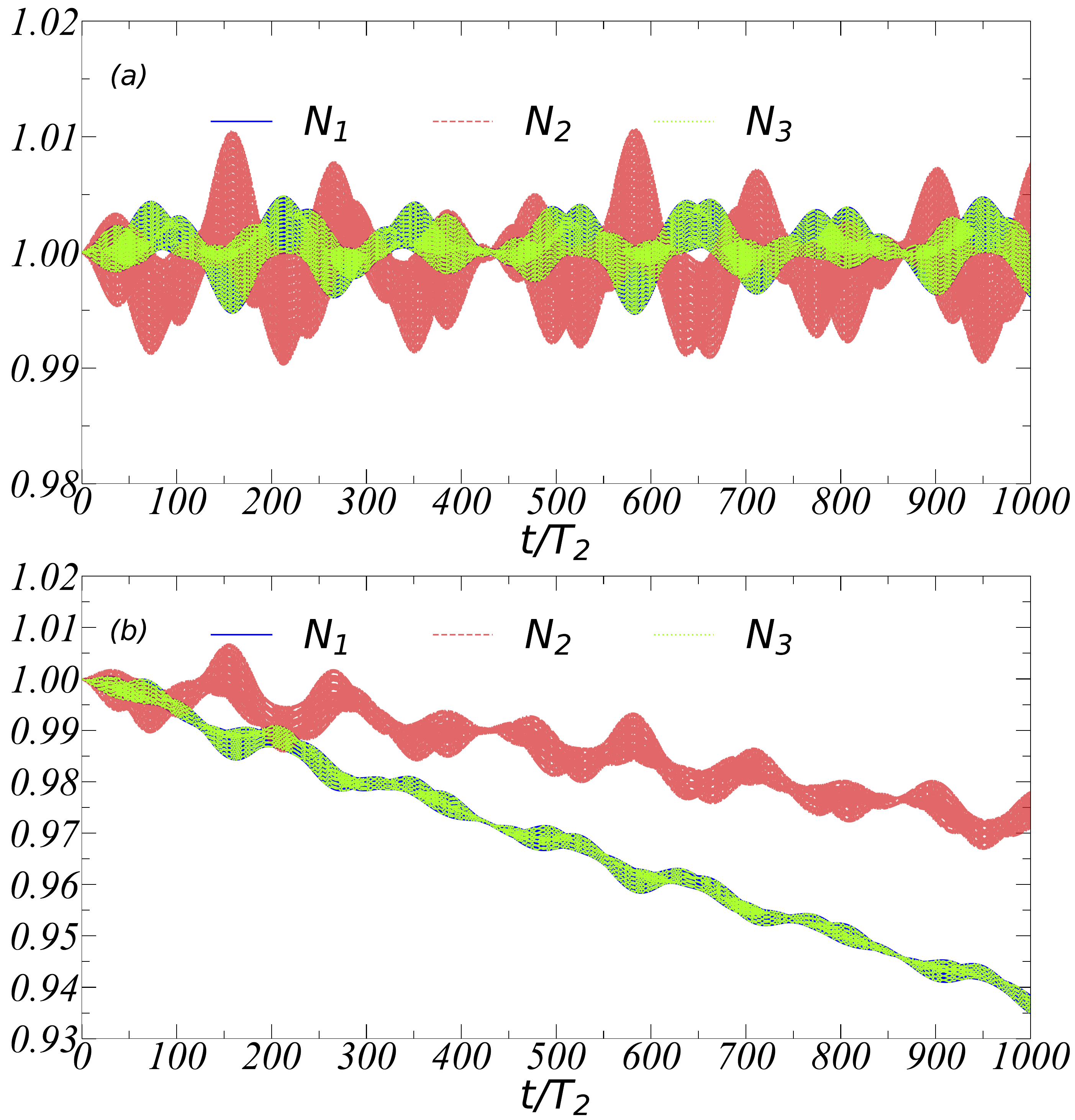}
\caption{We plot the average occupation number $\langle N_j(t)\rangle $ for each lattice site, for the integer resonance $\Omega_2=U/2$. The upper panel (a) shows the closed system scenario. The lower panel (b) shows the open system scenario. As in previous numerical simulations, the three-site Bose-Hubbard lattice is initialized in the state $\ket{\psi_0}=\ket{111}$. We use realistic parameters  $\omega=2\pi \times 4.5$ GHz, $J_0=2\pi \times 11.5$ MHz, $\kappa_{10}=11.9$ kHz, $\kappa_{21}=24.39$ kHz, $\kappa_{32}=33.33$ kHz, $\gamma_{00}=13.89$ kHz, $\gamma_{11}=31.25$ kHz, $\gamma_{22}=83.33$ kHz. We consider up to $n_{\rm max}=3$ particles per site with a local Hilbert space dimension $\dim{\mathcal{H}_\ell}=4$.}
\label{Fig5}
\end{figure}

\subsection{Three-site Bose-Hubbard lattice}
Let us consider a three-site Bose-Hubbard lattice. We will present results about the stability of the emerging many-body dynamics when considering integer ($\Omega_1=U$) and fractional ($\Omega_2=U/2$) resonance conditions in an open system scenario. Our study involves the numerical solution of the Lindblad master equation (\ref{MEBH}) using the fourth-order Runge-Kutta algorithm. We consider a product state with unit filling $\ket{\psi_0}=\ket{111}$ as an initial condition. In a closed system scenario, the system will only populate states within the positive parity subspace $\ket{\psi_0}$, $\ket{\psi_1}=\frac{1}{\sqrt{2}}(\ket{120}+\ket{021})$, $\ket{\psi_2}=\frac{1}{\sqrt{2}}(\ket{102}+\ket{201})$, $\ket{\psi_3}=\frac{1}{\sqrt{2}}(\ket{210}+\ket{012})$, $\ket{\psi_4}=\ket{030}$, $\ket{\psi_5}=\frac{1}{\sqrt{2}}(\ket{300}+\ket{003})$ \cite{PhysRevB.106.064307}. In an open quantum system scenario, we expect that U(1) symmetry will no longer be preserved. Our model will consider identical loss mechanisms for each lattice site, implying parity symmetry is still held.

In Figs. \ref{Fig2} and \ref{Fig3}, we plot the populations of states $\ket{\psi_i}$ with $i= 0, 1, 2, 3, 4, 5$ for the integer resonance condition $\Omega_1=U$ and fractional resonance condition $\Omega_2=U/2$, respectively. We identify populations as $P_i(t)=|\langle \psi_i|\psi(t)\rangle|^2$. In both figures, $P_i$ are populations numerically computed considering a closed system scenario governed by the Hamiltonian (\ref{eq:HBH}), and $P^D_i$ are populations numerically obtained considering decay and dephasing mechanisms acting upon each lattice site via the master equation (\ref{MEBH}). We stress that the populations of states $\ket{\psi_1}$ and $\ket{\psi_2}$ are the same, so we only show $P_1(t)$.

In our numerical simulations, we use realistic values for decay ($\kappa_{nm}$) and dephasing ($\gamma_{nn}$) rates of superconducting circuit experiments \cite{PhysRevLett.114.010501}. Also, single site frequency $\omega$ and hopping rate $J_0$ are taken from Ref.\cite{PhysRevResearch.3.033043}, and $U=40J_0$, see the caption of Fig.~\ref{Fig2} and Fig.~\ref{Fig3} for details. It is worth mentioning that despite the long time needed for the population of state $\ket{\psi_3}$ to occur with the highest probability in the fractional resonance case, see Fig.~\ref{Fig3}(b), the system is robust under loss mechanisms. Using realistic parameters, we estimate driving periods of about $T_1\approx 2$ ns and $T_2\approx 4$ ns, for integer and fractional resonance conditions, respectively. Although relaxation and dephasing affect the populations within the positive parity subspace, strong oscillations of probabilities still survive within our simulating time. This is a signature of the stability of many-body resonances. Strong oscillations still survive in local observables, such as the average occupation number per site. Figure \ref{Fig4} shows the average occupation number of each lattice site $N_j(t)$ for the integer resonance $\Omega_1=U$. The upper panel shows the dynamic evolution of the closed system, and the lower panel shows the dynamics in the open system scenario. As expected, $N_1(t)=N_3(t)$ due to the reflection symmetry of the lattice with identical parameters for loss mechanisms. Within the simulating time $t= 8~\mu$s ($t= 4000T_1)$, we still see strong oscillations of the occupation numbers and Poincare recurrences. Since we consider zero-temperature baths for relaxation, we expect $N_j(t)\to 0$ in the long-time dynamics. We stress that due to the memory consumption of the Runge-Kutta algorithm, we can not simulate longer times using the realistic parameters of Ref.\cite{PhysRevLett.114.010501}.

\begin{figure}[t]
\centering
\includegraphics[scale=0.4]{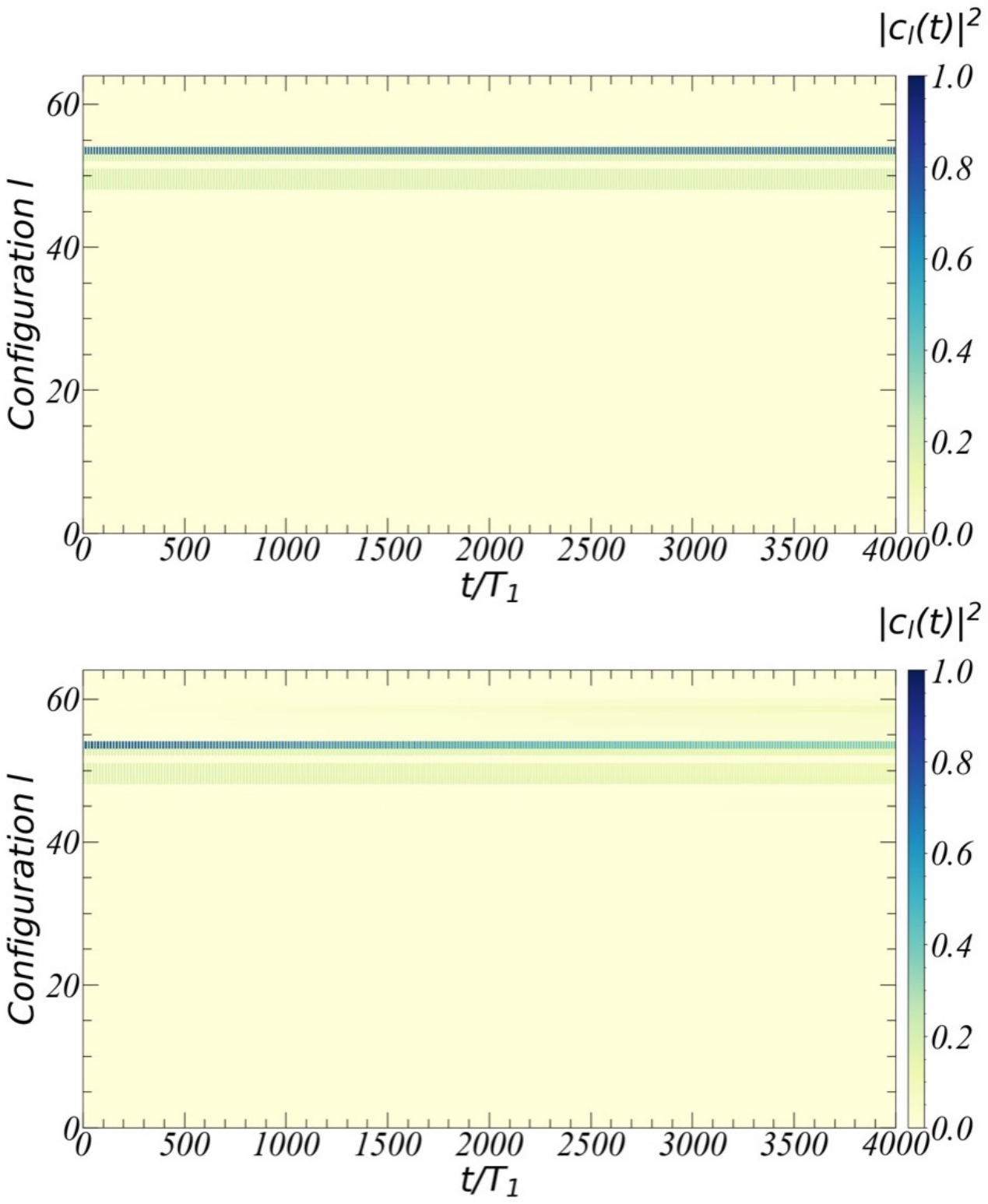}
\caption{Populations associated to each configuration $|c_{l}(t)|^2$ for the integer case $\Omega_1=U$, with $T_1=2\pi/\Omega_1$, and a lattice with $L=3$ sites. The upper panel shows the system dynamics considering a closed scenario, whereas the lower panel is an open scenario. In this simulation, we have considered all possible configurations in the Hilbert space, which contains $\mathcal{M}=64$ configurations. As in previous numerical simulations, we consider up to $n_{\rm max}=3$ particles per site with a local Hilbert space dimension $\dim{\mathcal{H}_\ell}=4$.}
\label{Fig6}
\end{figure}

\begin{figure}[t]
\centering
\includegraphics[scale=0.4]{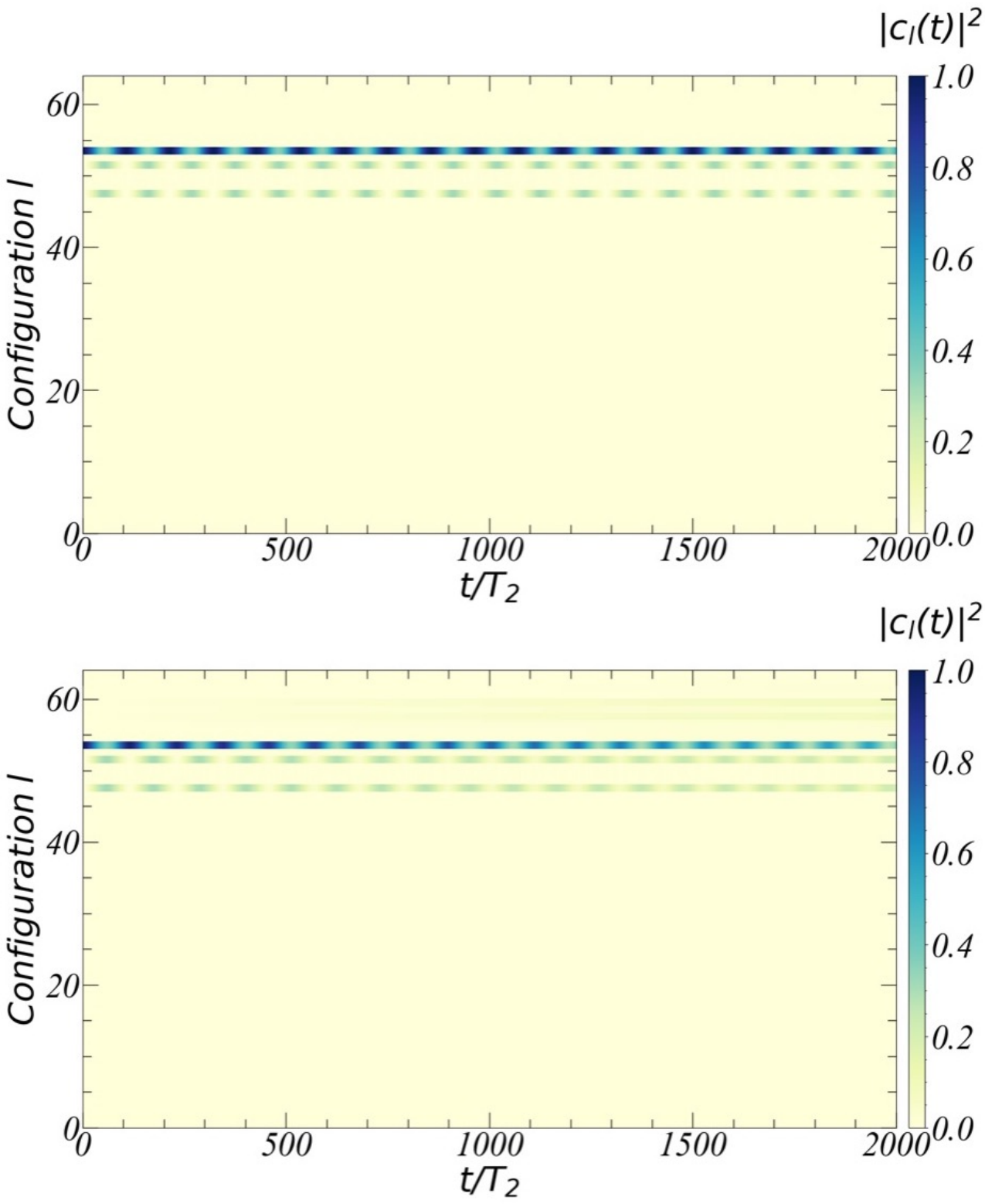}
\caption{Populations associated to each configuration $|c_{l}(t)|^2$ for the fractional case $\Omega_2=U/2$, with $T_2=2\pi/\Omega_2$, and a lattice with $L=3$ sites. The upper panel shows the system dynamics considering a closed scenario, whereas the lower panel is an open scenario. In this simulation, we have considered all possible configurations in the Hilbert space, which contains $\mathcal{M}=64$ configurations. As in previous numerical simulations, we consider up to $n_{\rm max}=3$ particles per site with a local Hilbert space dimension $\dim{\mathcal{H}_\ell}=4$.}
\label{Fig7}
\end{figure}
Figure \ref{Fig5} shows the average occupation number of each lattice site $N_j(t)$ for the fractional resonance $\Omega_2=U/2$. The upper panel shows the dynamic evolution of the closed system, and the lower panel shows the dynamics in the open system scenario. Although the average occupation numbers start decaying, within the simulating time $t= 4~\mu$s ($t= 1000T_2$), strong oscillations around the initial value $N_j(0)=1$ are still present. As in the integer resonance, we expect $N_j(t)\to 0$ in the long-time dynamics because zero-temperature baths act upon individual bosonic particles. Notice that $N_1(t)$ and $N_3(t)$ decay faster in comparison to $N_2(t)$. This happens because second-order processes dominate the dynamics, and lattice edges can be populated with more than one particle, which results in faster relaxation processes.

When considering noisy dynamics, the U(1) symmetry is broken, and the system could populate states outside the unit-filling subspace. In this case, the wave function may be written as a linear combination of all possible configurations $\ket{\psi(t)}=\sum_{l=1}^{\mathcal{M}}c_{l}(t)\ket{l}$, with $\mathcal{M}=\sum_{N=0}^LD_{N,L}$ for a fixed number of sites $L$. Therefore, it is necessary to compute the probability of all accessible configurations the system may visit along with its dynamical evolution. In the three-site Bose-Hubbard lattice, there are $\mathcal{M}=\sum_{N=0}^3D_{N,3}=64$ accessible configurations. Figure \ref{Fig6} shows the distribution of populations of each configuration $|c_{l}(t)|^2$ for the integer case $\Omega_1=U$. The upper panel shows the system dynamics considering a closed system, whereas the lower panel is an open system scenario. In the long-time dynamics, a fraction of configurations outside the unit filling subspace start to be populated; however, signatures of the stability of the integer resonance are visible in the average occupation number shown in Fig.\ref{Fig4}. In analogy, Fig.~\ref{Fig7} shows the distribution of populations of each configuration $|c_{l}(t)|^2$ for the fractional case $\Omega_2=U/2$. The upper panel shows the system dynamics considering a closed system, whereas the lower panel is an open system scenario.

\begin{figure}[t]
\centering
\includegraphics[scale=0.4]{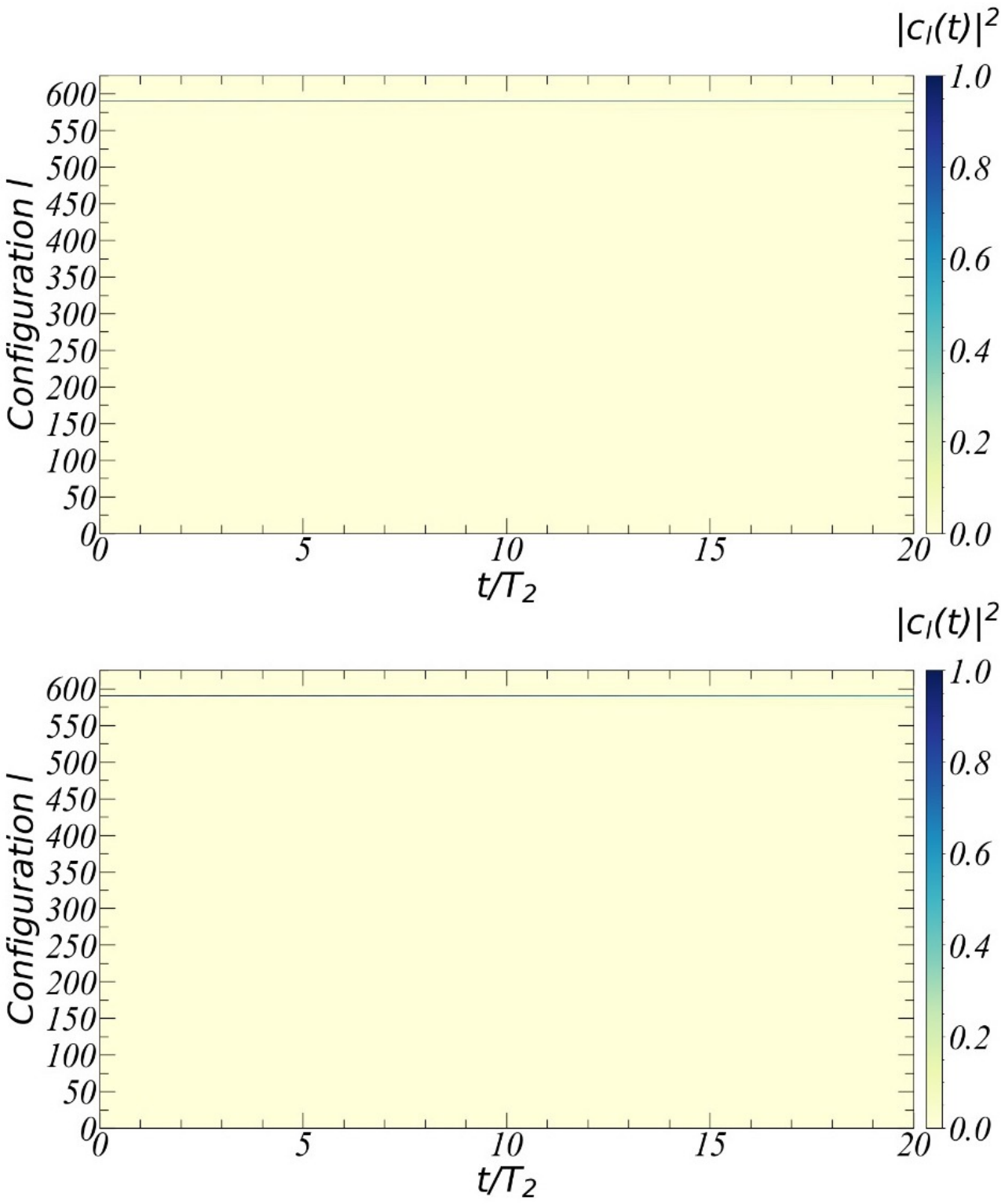}
\caption{Populations associated to each configuration $|c_{l}(t)|^2$ for the fractional case $\Omega_2=U/2$, with $T_2=2\pi/\Omega_2$, and a lattice with $L=4$ sites. The upper panel shows the system dynamics considering a closed scenario, whereas the lower panel is an open scenario. In this simulation, we have considered all possible configurations in the Hilbert space, which contains $\mathcal{M}=625$ configurations. In this case, we consider realistic parameters $\omega=2\pi \times 4.5$ GHz, $J_0=2\pi \times 11.5$ MHz, $\kappa_{10}=11.9$ kHz, $\kappa_{21}=24.39$ kHz, $\kappa_{32}=33.33$ kHz, $\kappa_{43}=500$ kHz, $\gamma_{00}=13.89$ kHz, $\gamma_{11}=31.25$ kHz, $\gamma_{22}=83.33$ kHz, and $\gamma_{33}=45.45$ kHz. We consider up to $n_{\rm max}=3$ particles per site with a local Hilbert space dimension $\dim{\mathcal{H}_\ell}=4$.}
\label{Fig8}
\end{figure}
\subsection{Four-site Bose-Hubbard lattice}
Let us consider the four-site Bose-Hubbard lattice initialized in the state $\ket{\psi_0}=\ket{1111}$. In this case, there are $\mathcal{M}=\sum_{N=0}^4D_{N,4}=625$ accessible configurations. As in the three-site Bose-Hubbard lattice, in Fig.~\ref{Fig8}, we plot the distribution of populations of each configuration $|c_{l}(t)|^2$ for the fractional case $\Omega_2=U/2$. The upper panel shows the system dynamics considering a closed system, whereas the lower panel is an open system scenario. In both cases, we see that the system populates the same configurations. Therefore, we conclude that, for finite lattice sites, and for short times up to $t=20~T_2$ (t=80 $\mu$s), the slowing down characteristic of the fractional resonance is a stable phenomenon under noisy dynamics. We emphasize that due to resource consumption of the Rung-Kutta algorithm and our limited computational resources, and our current computational resources, we can only simulate up to $t=20~T_2$ for a lattice of $L=4$ sites.
\begin{figure}[t]
\centering
\includegraphics[scale=0.20]{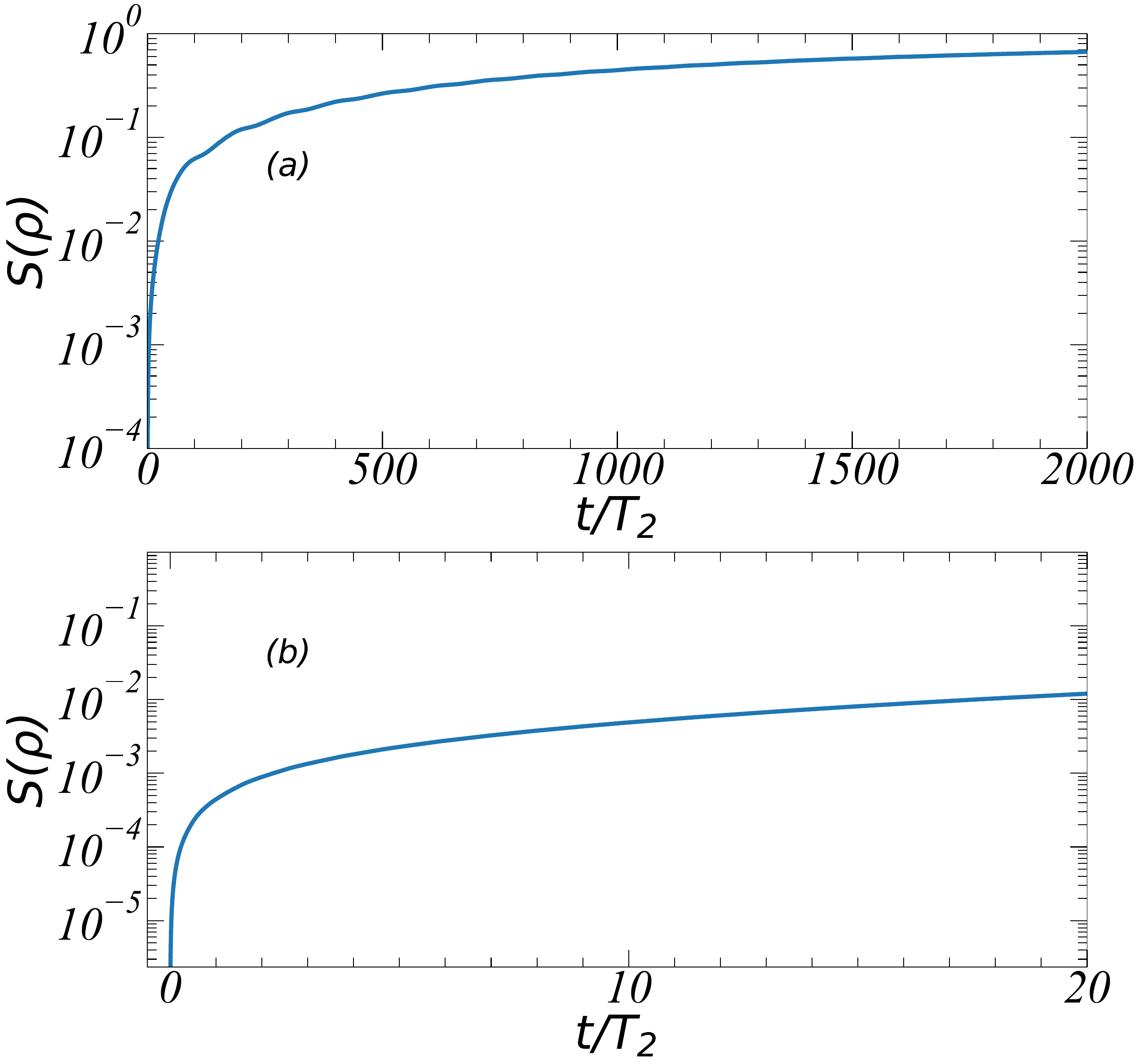}
\caption{Linear entropy for the fractional case $\Omega_2=U/2$, with $T_2=2\pi/\Omega_2$. (a) The upper panel represents a lattice of $L=3$ sites. (b) The lower panel shows a lattice of $L=4$ sites. We have used the same realistic parameters and local Hilbert space dimension as in previous numerical simulations.}
\label{Fig9}
\end{figure}

\subsection{The linear entropy}
Another way to characterize the stability of the system under noisy mechanisms is using the linear entropy defined as $S(\hat{\rho})=1-\rm{tr}(\hat{\rho}^2)$, where $\hat{\rho}$ is the system density matrix. In a closed system scenario, the linear entropy $S(\hat{\rho})=0$ at all times since $\hat{\rho}$ is a pure state. However, a realistic situation will necessarily imply the density matrix to be a statistical mixture. In the previous section, we demonstrated that the fractional many-body resonance and its characteristic slowing down of the dynamics \cite{PhysRevB.106.064307} is a stable phenomenon under loss mechanisms. The latter is also reflected in linear entropy as shown in figure \ref{Fig9}, where the upper (lower) panel shows the linear entropy as a function of time for a lattice of $L=3$ ($L=4$) sites. In both cases, $S(\hat{\rho})$ is is less than $1$ within the simulating time.

\section{Conclusions}
\label{conclusion}
Considering small lattices of strongly interacting bosonic particles, we have provided robust evidence about the stability of many-body resonances and their characteristic slowing down under realistic parameters of NISQ devices implemented in superconducting circuits. Our investigation includes decay and dephasing mechanisms acting locally on each bosonic particle using the Lindblad master equation, which has been proven helpful in describing state-of-the-art superconducting circuit experiments. In the short term, we expect to extend our study to larger lattice sizes $L>10$ using the adaptive time-dependent density matrix renormalization group approach to confirm further the findings presented here. Our conclusions are essential before seeking potential fractional resonance applications and their associated prethermal states as a quantum memory device.

\section*{Acknowledgments}
We thank R. Rom\'an-Ancheyta for helpful discussions and for carefully reading our article. R.P. acknowledges the support from Vicerrector\'ia de Postgrado USACH. G.R acknowledges the support from the Fondo Nacional de Desarrollo Cient\'ifico y Tecnol\'ogico (FONDECYT, Chile) Grant No. 1190727. We used time-evolving block decimation (TEBD) algorithm \cite{Vidal2004Jul,Hauschild2018Oct}, from \href{https://github.com/tenpy/tenpy}{https://github.com/tenpy/tenpy}.

\section*{Appendix: stability of integer resonance}

\begin{figure}[t]
\centering
\includegraphics[scale=0.1]{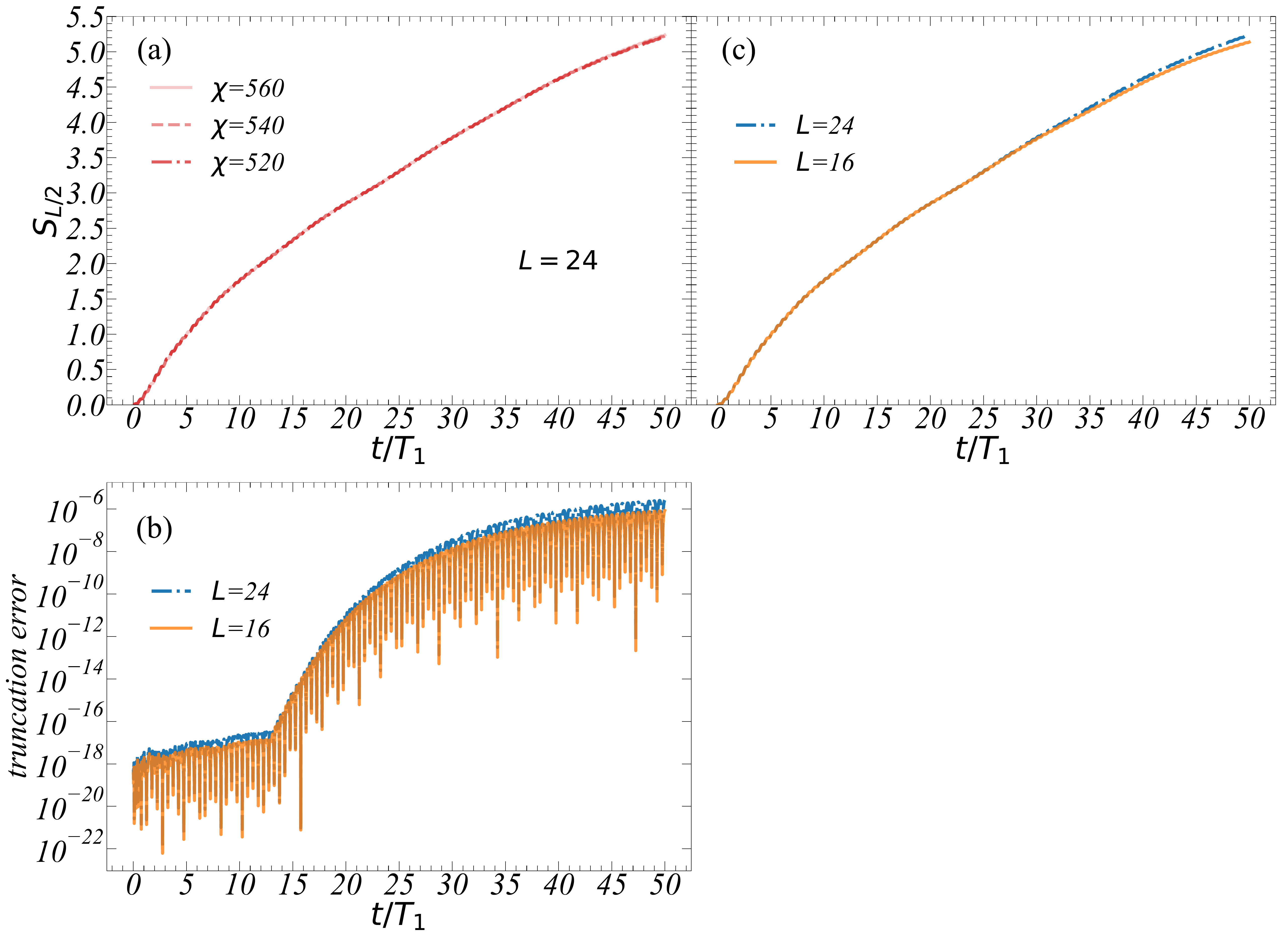}
\caption{(a) Convergence check of the TEBD algorithm by computing the half-chain von Neumann entropy, $S_{L/2}$, for a lattice of $L=24$ sites. (b) $S_{L/2}$ for lattices sizes $L=16$ and $L=24$. (c) Truncation errors of the TEBD algorithm for lattice sizes $L=16$ and $L=24$. In all numerical simulations, we considered parameters $J_0=0.01\omega$, $U=40J_0$, and up to $n_{\rm max}=3$ particles per site with a local Hilbert space dimension $\dim{\mathcal{H}_\ell}=4$.  }
\label{Fig10}
\end{figure}

In Ref.~\cite{PhysRevB.106.064307}, it has recently been shown that in the fractional resonance case, the half-chain entanglement entropy does not blow up within the simulating time as one increases the lattice size. The latter is due to the strong localization of the many-body state that results from the resonance condition. However, the stability of the integer resonance is not clear by simply comparing the entanglement behavior for two relatively nearby lengths of the lattice. Here, we have extended the lattice size up to $L=24$ sites, investigated the integer resonance dynamics, and considered a closed system scenario. Fig.\ref{Fig10}(a) shows the convergence check of the time-evolving block decimation algorithm (TEBD). The results show that a bond dimension $\chi=560$ is enough to ensure a trustful numerical simulation within the simulating time. The latter is further confirmed in Fig.\ref{Fig10}(b), where we plot the truncation error in a semi-log scale as a function of time. In this case, we compare the truncation errors for lattice sizes $L=16$ and $L=24$. In Fig.\ref{Fig10}(c), we compare $S_{L/2}$ for both lattice sizes. Our results show that by increasing the lattice size from $L=16$ to $L=24$, the half-chain von Neumann entropy does not blow up within the simulating time. The latter results from the strong localization of the quantum many-body state due to the integer resonance condition.

\bibliographystyle{apsrev4-1}
\bibliography{Mybib}
\end{document}